\documentclass[12pt,a4paper]{article}
\usepackage{amsmath}
\usepackage[dvips]{graphicx}
\input epsf
\usepackage{times}
\begin{document}
\begin{titlepage}
\title{Collective effects in multiparticle  production processes at the LHC}
\author{S. M. Troshin,
 N. E. Tyurin\\[1ex]
\small  \it Institute for High Energy Physics,\\
\small  \it Protvino, Moscow Region, 142281, Russia}
\normalsize
\date{}
\maketitle

\begin{abstract}
 We discuss various aspects of the multiparticle production processes at the LHC energy range with emphasis on
the collective effects associated with appearance of the new scattering mode, which corresponds
to the reflective scattering  and its impact on multiparticle production processes. 
 \\[2ex]
\end{abstract}
\end{titlepage}
\setcounter{page}{2}

\section*{Introduction}

Nowadays the LHC  is collecting data and providing  experimental results at the largest world
energy $\sqrt{s}=7$ $TeV$. 
Along with realization of its discovery potential,  the LHC experimental program  renewed interest to the well
known unsolved problems  providing deepened insights into those issues.
In this context the multiparticle production studies  bring us  a clue
to the mechanisms of confinement and hadronization. Confinement of a color (i.e. the fact that an isolated color object has an infinite energy in the physical vacuum)
is associated with collective, coherent interactions of quarks and gluons,
and results in formation of the asymptotic states, which are the
colorless, experimentally observable particles. 
The inelastic processes involve  large number
of particles in the final state. On  the other side, the experimental measurements often reveal high degrees of a coherence 
 in the relevant observables. No doubt these collective effects are very important for understanding of the nonperturbative collision dynamics.
 Such collective effects are associated, in particular, with
unitarity regulating the relative strength
of elastic and inelastic processes and connecting the amplitudes of the various multiparticle production processes.
Unitarity  for the scattering matrix is formulated
for  the asymptotic colorless  on-mass shell states. 
There is no  universal, generally accepted
method to implement full unitarity 
in high energy scattering.  A related problem of the absorptive
corrections  and of their sign
has a long history of discussion (cf. \cite{sachbla} and references therein).
However, a choice of particular unitarization scheme is not just a matter
of taste. 

Long time ago the  arguments based on analytical properties of the scattering
amplitude  were put forward \cite{blan} in favor of the rational form of unitarization.
It was shown  that this form  reproduced
correct analytical properties of the scattering amplitude
in the complex energy plane much easier compared to the
exponential form.  This specific form of unitarization \cite{umat} can be 
related to the confinement of color in QCD  \cite{umatn}. 

Correct account for the unitarity is  essential for the minimum bias multiparticle production processes,
correlations, anisotropic flows and studies of phase transitions. 

The important point is that the region of the LHC energies is the one where the new, reflective scattering
 mode \cite{reflect} can be observed. Such a mode naturally appears
 when energy grows and the rational form of unitarization
 being exploited \cite{phl93}.  
 This mode can be revealed at the LHC directly measuring
 $\sigma_{el}(s)$ and $\sigma_{tot}(s)$ \cite{reltot}.
 The reflective scattering is correlated with the self--damping of the inelastic channels and leads to the
 asymptotically dominating role of elastic scattering, i. e.
 ${\sigma_{el}(s)}/{\sigma_{tot}(s)}\rightarrow 1$ at $s\to\infty$ while the both cross-sections, $\sigma_{el}(s)$ and $\sigma_{tot}(s)$,
 tend to infinity in this
limit.

Of course, the obvious  questions appear  on compatibility of the reflective scattering mode with 
dynamics of the multiparticle production, in particular, with 
the growth of the mean multiplicity in hadronic collisions with energy.
Many models and the experimental data suggest a power-ilke energy dependence
of mean multiplicity\footnote{Discussions of power--like energy dependence
of  mean hadronic multiplicity and list of references to the older papers
can be found in \cite{polyak,barshay,menon}}. 

In this  review we discuss relation of the reflective scattering with confinement and hadronization, in particular, we consider 
the effects of the reflective scattering in the multiparticle
production processes at the LHC energies and apply
a rational ($U$--matrix)  unitarization
method \cite{umat}  to   consider  correlations in the
multiparticle dynamics.  We perform model calculations of global characteristics in the
multiparticle production processes such as mean multiplicity, average transverse momentum, two-particle correlations, elliptic flow
 and demonstrate its quantitative and qualitative 
agreement with the first  data
obtained at the LHC. It appeared that reflective scattering leads to the prominent effects in the global observables and particle
correlations.
\section{Rational form of unitarization and  confinement of colored degrees of freedom}

Unitarity or conservation of probability, which can be   written in terms
of the scattering matrix as
\begin{equation}\label{ss}
SS^+= \mathbf{1},
\end{equation} implies an
existence of the two scattering modes at high energies - shadowing and reflection.
Saturation of the
unitarity condition for  the scattering matrix
 in hadron collisions at small impact parameters $b$, takes place
when scattering acquires  reflective nature, i.e. $S(s,b)|_{b=0}\to -1$ at  $s\to\infty$.
Here $S(s,b)$ is the elastic scattering $S$-matrix in the impact parameter representation.
Approach  to the full absorption in head-on collisions ---
 the limit $S(s,b)|_{b=0}\to 0$ at $s\to\infty$ --- does not follow
from unitarity itself and is merely  a result of the assumed
saturation of the black disk limit.
On the other hand, the reflective scattering
is a natural interpretation of the unitarity saturation  based on the optical concepts in
high energy hadron scattering.
 Such reflective scattering can be traced to the
continuous  increasing density of
the scatterer with energy, i.e. when density goes beyond some critical value relevant
for the black disk limit saturation,
 the scatterer starts to
acquire  a reflective ability. The concept  itself is quite general,
 and results from the $S$-matrix  unitarity saturation
related to  the necessity to provide the total cross section
growth at $s\to\infty$. This picture predicts
  that the scattering amplitude at
  the LHC energies is beyond the black disk limit at small impact
  parameters.  The consequences for the total, elastic and
  inelastic cross-sections have been discussed in \cite{reltot}.

Unitarity  for the elastic scattering amplitude $F(s,t)$
  can be written in the form
 \begin{equation}\label{un}
 \mbox{Im}F(s,t)=H_{el} (s,t)+H_{inel} (s,t),
 \end{equation}
 where $H_{el,inel}(s,t)$ are the corresponding elastic  and inelastic overlap functions
 introduced by Van Hove \cite{vanh}. Physical meaning of each term in Eq. (\ref{un})
  is evident from the graphical representation in Fig. 1.
\begin{figure}[hbt]
\begin{center}
\includegraphics[scale=0.6]{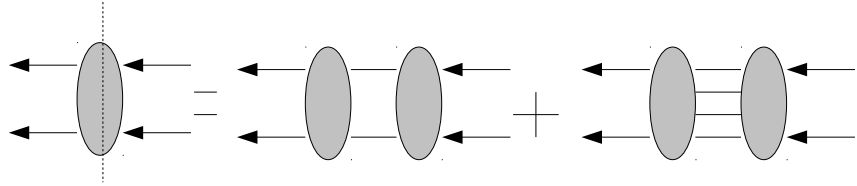}
\end{center}
\caption{\small\it Eq. \ref{un} in the graphical form.}
\end{figure}
The functions $H_{el,inel}(s,t)$ are related to the functions
$h_{el,inel}(s,b)$  via the Fourier-Bessel transforms, i.e.
\begin{equation}\label{hel}
H_{el,inel} (s,t)=\frac{s}{\pi^2}\int_{0}^{\infty} bdb h_{el,inel}(s,b) J_0(b\sqrt{-t}).
\end{equation}
The elastic and inelastic cross--sections can be obtained as
follows:
\begin{equation}\label{selin}
\sigma_{el,inel}(s)\sim \frac{1}{s} H_{el,inel} (s,t=0).
\end{equation}
As it was already noted, the reflective scattering mode appears
naturally
 in the $U$--matrix form of unitarization,
where the $2\to 2$ scattering matrix element in the
impact parameter representation
is the following linear fractional transform:
\begin{equation}
S(s,b)=\frac{1+iU(s,b)}{1-iU(s,b)}. \label{um}
\end{equation}
 $U(s,b)$ is the generalized reaction matrix, which is considered to be an
input dynamical quantity. The relation (\ref{um}) is one-to-one transform and easily
invertible.
 
Now, we would like to discuss relation of the rational unitarization form of scattering matrix with the confinement property of QCD.
According to the confinement, isolated colored objects cannot exist in the physical vacuum and  there is no room for objects like quark-proton scattering amplitude, 
since isolated color object has an infinite energy in the physical vacuum.

The important assumption in the derivation of the consequences of unitarity is the completeness of a set of the asymptotic states, those states include
 colorless degrees of freedom or hadrons only.
It can be
considered as a questionable one in the QCD era. It might be reasonable to claim that a set of hadronic states
does not provide a complete set of states and unitarity in the sense discussed above could be violated (cf. \cite{rlst}).
It was stated in this paper that the Hilbert space, which corresponds to colorless hadron states and is constructed using vectors
  spanned on the physical vacuum, should, in principle, be extended. 
At the present time it is  often underlined that the vacuum state  is not
unique; i.e. the colored  current quarks and gluons
are in fact the degrees of freedom related to a different vacuum. Thus, the vacuum state may not be considered anymore to be the state
of the lowest energy  (ground state).

 At the moment we  would like to demonstrate that inclusion of the states corresponding to the confined 
objects (e.g. colored current or constituent quarks) into a set of the asymptotic states  would lead to a rational form of S-matrix unitarization provided those states
satisfy a certain constraint treated 
as a condition for confinement. We propose to manage these states similar to consideration of the states with indefinite metric in quantum electrodynamics
as it was  performed by N.N. Bogolyubov \cite{nnb}.  
To construct the $S$ matrix (which operates in the physical subspace) let us consider state vectors  $|\Phi_-\rangle$ at $t \to - \infty$
and   $|\Phi_+\rangle$ at $t \to + \infty$ each being the sum of the two vectors
\[
 |\Phi_+ \rangle=|\varphi_+\rangle+|\omega_+\rangle
\]
\[
 |\Phi_- \rangle=|\varphi_-\rangle+|\omega_-\rangle
\]
where $|\varphi_\pm\rangle$ corresponds to the physical states and $|\omega_\pm\rangle$ -- 
to the confined states.  So, we  have that $|\varphi_\pm\rangle={\cal P}_\pm|\Phi_\pm\rangle$ 
and $|\omega_\pm\rangle=(1-{\cal P}_\pm)|\Phi_\pm\rangle$, 
where ${\cal P}_\pm$ are the relevant projection operators relevant for the initial and final states, respectively.
 
The scattering operator $\cal \tilde S$ (defined as $|\Phi_+\rangle={\cal \tilde S}|\Phi_-\rangle$)
should not, in principle, conserve probability and obey unitarity condition
since it operates in the  Hilbert space which includes subspace where confined objects  with
an undefined norm reside.  
Next, let us to impose condition on the asymptotic vectors  $|\omega_\pm\rangle$:
\[
| \omega_-\rangle+|\omega_+\rangle=0.
\]
 It means that {\it in-} and {\it out-} vectors corresponding to the states of the confined objects are just the mirror reflections of each other.
 Those reflections can be associated with the impossibility for confined objects to propagate outside the hadron. In quantum mechanics
such solution corresponds to standing  or stationary wave where on average  net propagation of energy is absent .
Thus, the rational form of unitary scattering operator $\cal S$ 
\[
 {\cal S}={\cal P}_+{\cal \tilde S}[1+(1-{\cal P}_+{\cal \tilde S})]^{-1}\equiv(1-{\cal U})(1+{\cal U})^{-1},
\]
in the physical subspace, i.e. $|\varphi_+\rangle={\cal S}|\varphi_-\rangle$)
can easily be obtained, since
\[
|\varphi_+\rangle={\cal P}_+{\cal \tilde S}( |\varphi_+\rangle+ |\omega_-\rangle).
\]

We have started with scattering operator ${\cal \tilde S}$ in the Hilbert space which includes
physical and unphysical subspaces and arrived to an
unitary scattering operator ${\cal  S}$ in the physical subspace.
Crucial assumption  was a constraint for the states of confined objects  from the unphysical subspace 
$| \omega_-\rangle+|\omega_+\rangle=0$, which
we assume to be related to  the confinement condition. The imposition of this constraint is 
an equivalent to the statement that the scattering
matrix in the unphysical subspace is identical with -\bf 1\rm .
Thus, one can say that if scattering matrix in the  space spanned over confined states
 is identically equal to  -\bf 1\rm, then scattering
matrix   in the physical space is unitary.
It is plausible therefore to assume  that unitarity can   be straightforwardly connected to confinement.

Rational or $U$-matrix form of unitarization was proposed long time ago \cite{heit} in the theory of radiation dumping.
Self-damping of inelastic channels was considered in \cite{bbla} and
 for the  relativistic case such form of unitarization was obtained in \cite{umat}. 
It was demonstrated that it can be used for construction of a bridge
between the physical states of hadrons and the states of the confined colored objects. 

\section{Reflective scattering at the LHC energies}

In the rational form of unitarization the inelastic overlap function $h_{inel}(s,b)$
is connected with the function $U(s,b)$ by the relation
\begin{equation}\label{hiu}
h_{inel}(s,b)=\frac{\mbox{Im} U(s,b)}{|1-iU(s,b)|^{2}},
\end{equation}
and the only condition to obey unitarity
 is $\mbox{Im} U(s,b)\geq 0$. The elastic overlap function is related to the function
 $U(s,b)$ as follows
\begin{equation}\label{heu}
h_{el}(s,b)=\frac{|U(s,b)|^{2}}{|1-iU(s,b)|^{2}}.
\end{equation}
The form of $U(s,b)$ depends on  particular model assumptions,
but for  qualitative
 purposes it
is sufficient that it increases with energy in a power-like way
 and decreases with the impact parameter
like a linear exponent or Gaussian.
To simplify the qualitative picture, we consider also the function $U(s,b)$
as  a pure imaginary.
At sufficiently  high energies ($s>s_R$),
the two separate  regions of
 impact parameter distances can be anticipated, namely the outer region
of peripheral collisions where the scattering has a typical absorptive origin, i.e.
$S(s,b)|_{b>R(s)}>0$ and
 the inner region of central collisions
where the scattering has a combined reflective and absorptive origin, $S(s,b)|_{b< R(s)}<0$.

We discuss now the impact parameter
profiles of elastic and inelastic overlap functions for the scattering picture with reflection.
 The well known absorptive picture of high energy scattering corresponds to the black
 disk limit  at $s\to\infty$. This limit   has already been reached in the head-on
  proton--antiproton  collisions at Tevatron, i.e. $h_{el}(s,b=0)\simeq 0.25$ \cite{gir}. Thus, one can expect that
  in the framework of absorptive picture, black disk limit will be reached also at $b\neq 0$ at higher
  energies and the  profiles of $h_{el}(s,b)$ and $h_{inel}(s,b)$
  are to  be similar  and have a form close to  step function (Fig. 2).
\begin{figure}[hbt]
\begin{center}
\includegraphics[scale=0.6]{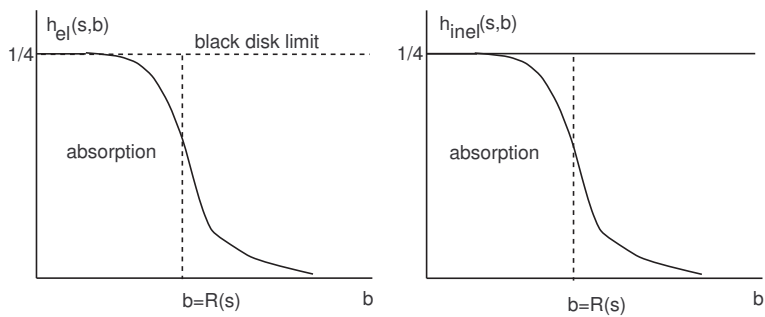}
\end{center}
\caption{\small\it Typical picture of impact parameter profiles of the
elastic and inelastic overlap function in the absorptive approach at asymptotic
 energies.}
\end{figure}
We consider now energy evolution of the elastic and inelastic overlap functions
 which includes reflective scattering.
 With
conventional parametrization of the $U$--matrix
 the inelastic overlap function increases with energies
at modest values of $s$. It reaches its maximum value $h_{inel}(s,b=0)=1/4$ at some
energy $s=s_R$ and beyond this energy the  reflective
scattering mode appears at small values of $b$. The region of energies and
impact parameters corresponding
to the reflective scattering mode is determined by the conditions
$h_{el}(s,b)> 1/4$ and $h_{inel}(s,b)< 1/4$. The unitarity limit and black disk
limit are the same for the inelastic overlap function, but these limits are different
for the elastic overlap function.
The quantitative analysis of the experimental data
 \cite{pras} gives the threshold value: $\sqrt{s_R}\simeq 2$ TeV.
The function $h_{inel}(s,b)$ becomes peripheral when energy increases
in the region $s>s_R$.
At such energies the inelastic overlap function reaches its maximum
 at $b=R(s)$ where  $R(s)\sim \ln s$.
So, in the energy region, which lies beyond the transition threshold,
 there are two regions in impact
 parameter space: the central region
of reflective scattering combined with absorptive scattering
at $b< R(s)$ and the peripheral region
of pure absorptive scattering at $b> R(s)$. Typical pattern of the impact
parameter profiles for elastic and inelastic overlap functions in the scattering
picture with reflection at the LHC energies is depicted on Fig. 3.
\begin{figure}[hbt]
\begin{center}
\includegraphics[scale=0.6]{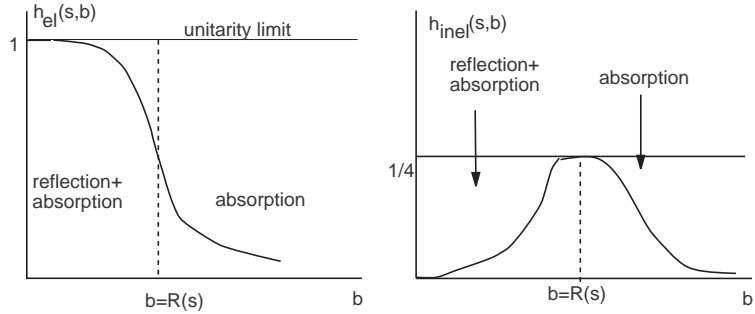}
\end{center}
\caption{\small\it Typical qualitative picture of impact parameter profiles of the
elastic and inelastic overlap function in the reflective approach at the
 asymptotically high energies.}
\end{figure}

It should be noted that at the  values of energy $s>s_R$ the equation $U(s,b)=1$ has a solution in the
physical region of impact parameter values, i.e. $S(s,b)=0$ at $b=R(s)$. This line is shown
in  the $s$ and $b$ plane in Fig. 4 alongside with the
regions where elastic $S$-matrix has positive and negative values.
   \begin{figure}[hbt]
\begin{center}
\includegraphics[scale=0.5]{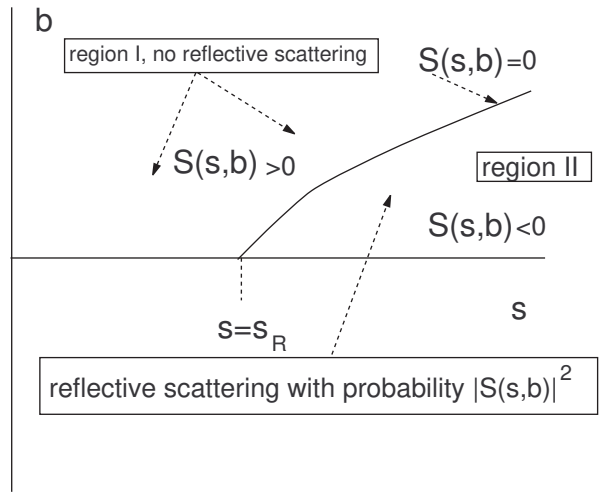}
\caption{\small\it Regions of positive (absorptive scattering) and negative 
(absorptive and reflective scattering) values of the
function $S(s,b)$ in the $s$ and $b$ plane.}
\end{center}
\end{figure}
The dependence of $S(s,b)$ on impact
 parameter $b$ at fixed energies (in the region $s>s_R$) is depicted on
 Fig. 5.
\begin{figure}[hbt]
\begin{center}
\includegraphics[scale=0.4]{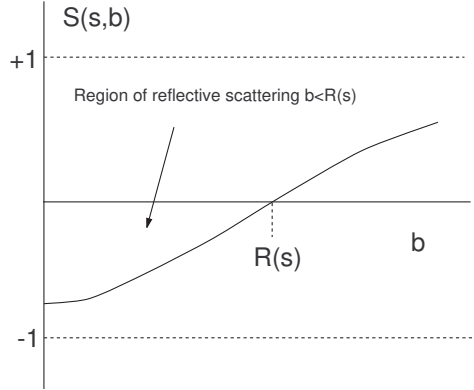}
\caption{\small\it Qualitative impact parameter dependence  of the
function $S(s,b)$ for  the energies  $s>s_R$.}
\end{center}
\end{figure}

The probability of reflective scattering at $b<R(s)$ and $s> s_R$ is determined by the magnitude
 of $|S(s,b)|^2$; this probability is equal to zero at $s\leq s_R$ and $b\geq R(s)$ (region I on Fig.4).
The behavior of $R(s)$
 is determined  by the logarithmic  dependence $R(s) \sim \frac{1}{M}\ln s$. It is
  consistent with the analytical properties of the resulting elastics scattering amplitude in
  the complex $t$-plane and mass $M$ can be related to the pion mass.

 Thus, at the energies $s> s_R$  the reflective scattering will mimic presence of repulsive core in
 hadron and meson interactions as well. The
generic geometrical
 picture at fixed energy beyond the black disc limit is described
as a scattering off
the partially reflective and partially absorptive disk
surrounded by the black ring which becomes gray at larger values of the
impact parameter.  The evolution with energy  is characterized
by increasing albedo due to the  interrelated  increase of reflection
  and decrease of absorption at small impact parameters.
Asymptotically, picture of  particle collisions with small impact parameters
looks like collisions of hard spheres.

\section{Multiparticle production in the $U$--matrix approach}
To consider multiparticle production in the $U$--matrix approach
it should be noted first that
\begin{equation}\label{imu}
\mbox{Im} U(s,b)=\sum_{n\geq 3} \bar U_n(s,b),
\end{equation}
where $\bar U_n(s,b)$ is a Fourier--Bessel transform of the function
\begin{eqnarray}\label{unn}
\bar U_n(s,t) & = & \frac{1}{n!}\int \prod_{i=1}^n\frac{d^3q_i}{q_{i0}}\delta^{(4)}(\sum_{i=1}^n q_i-p_a-p_b)
U_n^*(q_1,....,q_n;p_{a}',p_{b}')\cdot\\
& &U_n(q_1,....,q_n;p_a,p_b)\nonumber.
\end{eqnarray}
Here the functions $U_n(q_1,....,q_n;p_a,p_b)$ and $U_n(q_1,....,q_n;p_{a'},p_{b'})$ correspond to the
ununitarized (input or ``Born'') amplitudes of the process
\[
a+b\to 1+....+n,
\]
and the process with the same final state and the initial state with different momenta $p_a'$ and $p_b'$
\[
a'+b'\to 1+....+n,
\]
respectively. They are the analogs of the elastic $U$-matrix
for the processes $2\to n$. It is important to note that the functions $\bar U_n(s,t)$ are real ones (but not positively defined, contrary to the functions 
$\bar U_n(s,b)$). The impact parameter $b$ is the variable which is conjugated to the variable $\sqrt{-t}$, where
$t=(p_a-p_a')^2$ and  the following sum rule is valid for the function $I(s,b,q)$
\begin{equation}\label{sum}
\int \frac{d^3q}{E}I(s,b,q)=\langle n \rangle (s,b)\mbox{Im} U(s,b).
\end{equation}

The sum in the right hand side of the
Eq. (\ref{unn}) runs over all  inelastic final  states $|n\rangle$
 which include  diffractive  as
well as non-diffractive ones. Graphically, these relations are illustrated in Fig. 6.
\begin{figure}[hbt]
\begin{center}
\includegraphics[scale=0.5]{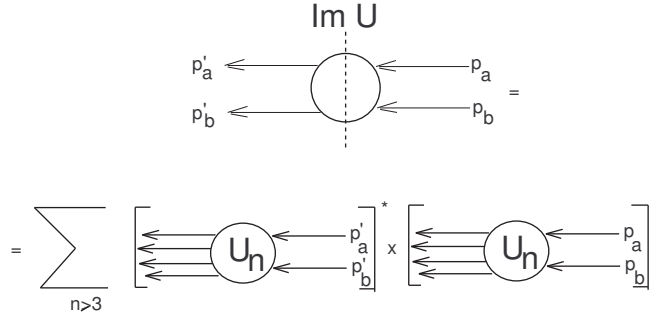}
\end{center}
\vspace*{8pt}
\caption{\small\it Unitarity  for the function $\mbox{Im} U$.}
\end{figure}

Then the inclusive cross-section of the process $ab\to cX$ has the following
form\cite{tmf,tmf1}, which is similar to the expression for the total inelastic
cross--section:
\begin{equation}
E\frac{d\sigma}{d^3q}= 8\pi\int_0^\infty
bdb\frac{I(s,b,q)}{|1-iU(s,b)|^2}\label{unp},
\end{equation}
where $I(s,b,q)$ is the Fourier-Bessel transform of the functions which are defined
similar to Eq. (\ref{unn}) but with the fixed momentum $q$ and energy $E$ of the particle $c$ in
 the final state (Fig. 7).
\begin{figure}[hbt]
\begin{center}
\includegraphics[scale=0.5]{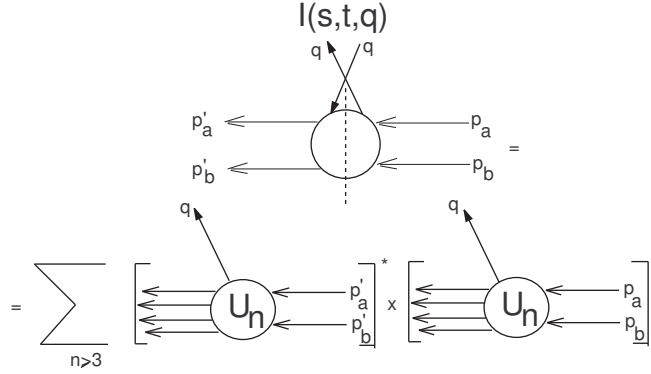}
\end{center}
\vspace*{8pt}
\caption{\small\it Unitarity  for the function $I(s,t,q)$.}
\end{figure}

The impact parameter $b$ is related to the impact parameters of the
secondary particles by the relation \cite{sakai}
\begin{equation}\label{v}
\mathbf{b}=\sum_{i=1}^n x_i\mathbf{b}_i,
\end{equation}
where $x_i$ stands for the Feynman variable $x$ of the $i$-th particle.

The general relations listed above reflect  the unitarity saturation only and do not
include other interaction dynamics. To provide further insight into the mechanism of multiparticle
production additional assumptions on the quark-gluon hadron structure and their interaction dynamics should 
be adopted. It is described in the following two sections

\section{Transient state of matter in hadron interactions}
We assume that the transient states of matter in hadron and nuclei collisions have the same nature   and originate from 
nonperturbative sector of QCD,   associated with
the mechanism of spontaneous chiral symmetry breaking ($\chi$SB) in QCD \cite{bjorken}. Due to this mechanism
transition of current into  constituent quarks takes place, the latter ones are  the quasiparticles whose masses are comparable with  a typical
 hadron mass scale.
 These constituent quarks interact via exchange
of the Goldstone bosons which  are collective excitations of the condensate and are represented by pions (cf. e.g. \cite{diak}).
The  general form of the effective Lagrangian (${\cal{L}}_{QCD}\rightarrow {\cal{L}}_{eff}$)
 relevant for description of the non--perturbative phase of QCD
 includes the three terms \cite{gold} \[
{\cal{L}}_{eff}={\cal{L}}_\chi +{\cal{L}}_I+{\cal{L}}_C.\label{ef} \]
Here ${\cal{L}}_\chi $ is  responsible for the spontaneous
chiral symmetry breaking and turns on first.  To account for the
constituent quark interaction and their confinement the terms ${\cal{L}}_I$
and ${\cal{L}}_C$ are introduced.  ${\cal{L}}_I$ and
${\cal{L}}_C$ do not affect the internal structure of the constituent
quarks.

The picture of a hadrons consisting of the constituent quarks embedded
 into quark condensate implies that overlapping and interaction of the
peripheral clouds   occurs at the first stage of hadron interaction.

\begin{figure}[htb]
\hspace{3cm}
\epsfxsize=3in \epsfysize=1.85in\epsffile{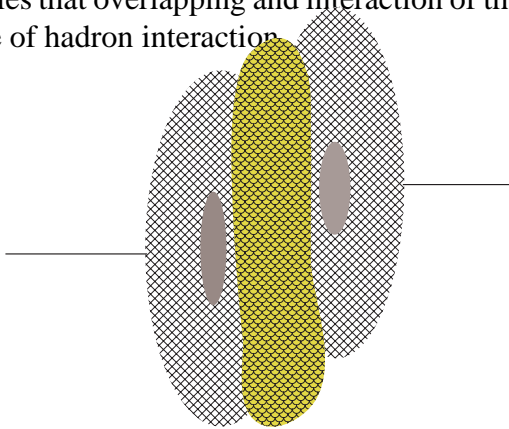}
 \caption[illyi]{{\small\it Schematic view of initial stage of the hadron
 interaction}.}
\label{ill5}
\end{figure}
Nonlinear field couplings   transform then some part of their kinetic energy into
internal energy related to a mass \cite{heis,carr}. 
As a result  we assume that the massive
virtual quarks appear in the overlapping region and  the effective
field is generated in that way. This field is generated by $\bar{Q}Q$ pairs and
pions strongly interacting with quarks. In their turn pions themselves are bound states of the constituent
quarks. At this stage the part of the effective Lagrangian ${\cal{L}}_C$ is turned off
(it turns on again in the final stage of the reaction) and  interaction is
 described by ${\cal{L}}_I$. Its
possible form  was discussed in \cite{diakp}.
The  transient phase (effective field)  generation time $\Delta t_{tp}$
\[
\Delta t_{tp}\ll \Delta t_{int},
\]
where $\Delta t_{int}$ is the total interaction time. This assumption on the almost instantaneous
generation of the effective field has obtained a support in the very short thermalization time revealed
in the heavy-ion collisions at RHIC \cite{therm,therm1}.

 This picture assumes  deconfinement at the initial stage of
 the hadron collisions and  generation of the effective field common for both hadrons.
Such ideas were  used in the model \cite{csn} which has
been applied to description of the elastic scattering. Massive virtual quarks play a role
of scatterers for the valence quarks in elastic scattering while
 their hadronization leads to
production of the secondary particles. 
 To estimate the number
of such scatterers one could assume that certain  part of hadron energy carried by
the outer condensate clouds is being released in the overlap region
 to generate the massive quarks. Then this number can be estimated  as
 \begin{equation} \tilde{N}(s,b)\,\propto
\,\frac{(1-\langle k_Q\rangle)\sqrt{s}}{m_Q}\;D^{h_1}_c\otimes D^{h_2}_c
\equiv N_0(s)D_C(b),
\label{Nsbt}
\end{equation} where $m_Q$ -- constituent quark mass, $\langle k_Q\rangle $ --
average fraction of hadron  energy carried  by  the constituent valence quarks. Here the function $D^h_c$
describes the condensate distribution inside the hadron $h$, and $b$ is an impact parameter of the colliding hadrons.
Thus, $\tilde{N}(s,b)$ massive quarks appear in addition to $N=n_{h_1}+n_{h_2}$
valence quarks. In elastic scattering those quarks are transient
ones: they are transformed back into the condensates of the final
hadrons. Calculation of the elastic scattering amplitude has been performed
in \cite{csn}.
However,  valence quarks can excite a part of the cloud of the virtual massive
quarks and those  will subsequently fragment into the multiparticle
final states. Such mechanism is responsible for the  particle production
and should lead to  correlations between
secondary particles which originate from the particular quark cluster resulting
from the valence quark excitation. In this impact parameter picture the strong forward--backward
multiplicity correlations should be expected. 

Another mechanism contributing to the multiparticle production is a direct  (i.e. not induced by interactions with valence
quarks) hadronization of the massive quarks.

\section{Multiparticle production mechanism}
In sections 5 and 6 we consider correlations arising in the average multiplicity 
and transverse momentum behavior.

Remarkably,  the existence of the massive quark-antiquark matter at the stage
preceding hadronization seems to be
supported  by the experimental data obtained
at CERN SPS and RHIC (see \cite{biro} and references therein).

Since the quarks are constituent, it is natural to expect  direct
proportionality between a secondary particles multiplicity  and
number of virtual massive quarks appeared in  collision of the  hadrons
with  given impact parameter:
\begin{equation}\label{mmult}
\langle n\rangle (s,b)=\alpha (n_{h_1}+n_{h_2})N_0(s)D_F(b)+ \beta N_0(s)D_C(b),
\end{equation}
with  constant factors $\alpha$ and $\beta$ and
\[
D_F(b)\equiv D_Q\otimes D_C,
\]
where the function $D_Q(b)$ is the probability amplitude of the interaction of
valence quark, which is in fact related
to the quark matter distribution in this hadron-like object  \cite{csn}.
The mean multiplicity $\langle n\rangle (s)$ can be calculated according to the
formula
\begin{equation}\label{mm}
\langle n\rangle (s)= \frac{\int_0^\infty  \langle n\rangle  (s,b)h_{inel}(s,b)bdb}{\int_0^\infty h_{inel}(s,b)bdb}.
\end{equation}
It is evident that the peripheral profile of $h_{inel}(s,b)$  associated with reflective scattering 
suppresses the region of small
impact parameters and the main contribution to the mean multiplicity is due to
the region of $b\sim R(s)$.

To make explicit calculations  we model 
the condensate distribution $D_C(b)$ and the impact parameter dependence
of the probability amplitude $D_Q(b)$
  by the exponential forms
with the
 different radii.
  Then the mean multiplicity
\begin{equation}\label{nsbex}
\langle n\rangle  (s,b)=\tilde\alpha N_0(s)\exp (-b/R_F)+\tilde\beta N_0(s)\exp (-b/R_C).
\end{equation}
The function $U(s,b)$   is
chosen as a product of the averaged quark amplitudes \begin{equation}
U(s,b) = \prod^{N}_{Q=1} \langle f_Q(s,b)\rangle. \end{equation}  
  This factorization originates
from an assumption of a  quasi-independent  nature  of the valence
quark scattering, $N$ is the total number of valence quarks in the colliding hadrons.
The $b$--dependence of  $\langle f_Q \rangle$ related to
 the quark formfactor $F_Q(q)$ has a simple form $\langle
f_Q\rangle\propto\exp(-m_Qb/\xi )$.
Thus, the generalized
reaction matrix (in a pure imaginary case) gets
the  form \cite{csn}
\begin{equation} U(s,b) = ig\left [1+\alpha
\frac{\sqrt{s}}{m_Q}\right]^N \exp(-Mb/\xi ), \label{x}
\end{equation} where $M =\sum^N_{q=1}m_Q$.
At sufficiently high energies where increase of the total cross--sections
is quite prominent  we
can neglect the energy independent term  and rewrite
the expression for $U(s,b)$ as  \begin{equation}
U(s,b)=i{g}\left(s/m^2_Q\right)^{N/2}\exp (-Mb/\xi ).
\label{xh} \end{equation}

After calculation of the integrals (\ref{mm})
 we arrive to the power-like dependence
of the mean multiplicity $\langle n\rangle (s)$ at high energies
\begin{equation}\label{asm}
\langle n\rangle (s) = as^{\delta_F}+bs^{\delta_C},
\end{equation}
where
\[
\delta_{F}={\frac{1}{2}\left(1-\frac{\xi }{m_QR_{F}}\right)}\quad
\mbox{and}\quad \delta_{C}={\frac{1}{2}\left(1-\frac{\xi }{m_QR_{C}}\right)}.
\]
There are four free model parameters, $\tilde\alpha$, $\tilde\beta$ and
$R_F$, $R_C$, and the freedom
in their choice is translated to  $a$, $b$ and
 $\delta_F$, $\delta_C$.
 The value of  $\xi=2$ is fixed from the data on angular
 distributions \cite{csn} and for the mass of constituent quark
 the standard value $m_Q=0.35$ GeV was taken. However, fit to experimental data on the
 mean multiplicity leads to approximate equality $\delta_F\simeq \delta_C$ and
actually Eq. (\ref{asm}) is reduced to the two-parametric power-like energy dependence
 of mean multiplicity
\[
 \langle n\rangle =as^\delta,
 \]
 which is in good agreement with the experimental data (Fig. 9). Equality
 $\delta_F\simeq \delta_C$  means that variation of the correlation strength with
 energy is weaker than the power dependence and could be, e.g. a logarithmic one.
  From the comparison
 with the data  on mean multiplicity we obtain that
$\delta\simeq 0.2$, which corresponds to the effective masses, which are determined
by the respective radii ($M=1/R$),
$M_C\simeq M_F\simeq 0.3m_Q$, i.e. $M_F\simeq M_C\simeq m_\pi$.
The value of mean multiplicity expected at the LHC energy
($\sqrt{s}=14$ TeV) is about 110.
 \begin{figure}[t]
\begin{center}
\includegraphics[scale=0.5]{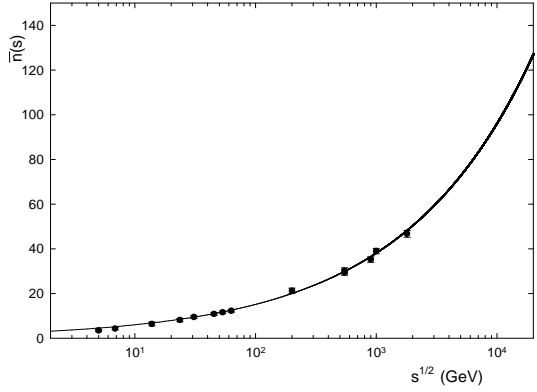}
\end{center}
\vspace*{8pt}
\caption{\small\it Energy dependence of mean multiplicity, theoretical curve
  is given by the equation $\langle n\rangle (s)=as^\delta$ ($a=2.328$, $\delta = 0.201$).}
\end{figure}

Now we would like to make a remark on the mean multiplicity  in the impact parameter representation. As it follows from the formulas
of the section 2,
the $n$--particle production
cross--section $\sigma_n(s,b)$
\begin{equation}\label{snb}
\sigma_n(s,b)=\frac{\bar{U}_n(s,b)}{|1-iU(s,b)|^2}
\end{equation}
Then the probability
\begin{equation}\label{pnb}
  P_n(s,b)\equiv\frac{\sigma_n(s,b)}{\sigma_{inel}(s,b)}=\frac{\bar{U}_n(s,b)}{\mbox{Im} U(s,b)}.
\end{equation}

Thus, we observe  cancellation of the unitarity corrections in
the ratio of the cross-sections $\sigma_n(s,b)$ and $\sigma_{inel}(s,b)$.
Therefore the mean multiplicity in the impact parameter representation
\[
\langle n\rangle  (s,b)=\sum_n nP_n(s,b)
\]
 is not affected by unitarity corrections. 
However, the above cancellation of unitarity corrections
 does not take place for the quantity $\langle n\rangle  (s)$. 
The main contribution  to $\langle n\rangle  (s)$ originates from 
peripheral values of the impact parameter. Thus, typical inelastic event
at the LHC energies is the event with a nonzero value of the initial impact 
parameter of collision.
\section{Rotation of transient matter and energy dependence of average transverse momentum}
We are going now to evaluate energy dependence of the average transverse momentum of produced particles
 and propose a possible
mechanism leading to this dependence. 
 The  geometrical picture of hadron collision discussed above
implies that at high energies and non-zero impact parameters
 the constituent quarks produced in overlap region carry  large orbital angular momentum.
 It  can be estimated
as 
\begin{equation}\label{l}
 L(s,b) \propto  b \frac{\sqrt{s}}{2}D_C(b).
\end{equation}
Due to supposed strong interaction
between the quarks this orbital angular momentum will lead to a coherent rotation
of the quark system located in the overlap region as a whole.
This rotation is similar to rotation of the liquid
where strong correlations between particles momenta exist \cite{qgpo}.
In what follows we argue that this collective coherent rotation would lead to the energy dependence of the
average transverse momentum and can explain experimentally observed rising behavior of this quantity.
It should be noted that 
discovery of the deconfined state of matter has
been announced  by four major experiments at RHIC \cite{rhic}.
Despite the highest values of energy and density have been reached,
a genuine quark-gluon plasma QGP (gas of the free current quarks and gluons)
was not observed. The deconfined state reveals the properties of the perfect liquid,
being a strongly interacting collective state and therefore it was labeled as sQGP.
Using similarity between the hadronic  and nuclear interactions we assumed that transient state in
hadron interactions is also a liquid-like strongly interacting matter. As it was already 
mentioned,  the presence of large angular momentum in the overlap region
will lead to coherent rotation of quark-pion liquid. Of course, there should be experimental
observations of this collective effect and one of them is the directed flow in hadron reactions,
with fixed impact parameter discussed in \cite{qgpo}. It is not impossible task to measure impact
parameter of collision in hadron reactions with the help of the event multiplicity studies. 
But effects averaged over impact parameter can be measured more easily using
standard experimental techniques. So, it is natural to assume that the rotation of transient matter
will affect average transverse momentum of the secondary hadrons in proton-proton collisions.
Let for beginning do not take into account  the other sources of the transverse momentum  and temporally
suppose that all average transverse momentum is a result of a coherent rotation of transient liquid-like state. 
Then the following relation can be invoked
\begin{equation} \label{ptl}
\langle p_T\rangle(s,b)=\kappa L(s,b),
\end{equation}
 where $L(s,b)$ is given by Eq. (\ref{l}) and $\kappa$ is a constant which has  dimension of inverse length.
It is natural to relate it with inverse hadron radius,  $\kappa\sim 1/R_h$.
To calculate the average transverse momentum $\langle p_T(s)\rangle$  the following relation
 with  $\langle p_T \rangle (s,b)$ will be used:
\begin{equation}\label{mpt}
 \langle p_T\rangle (s)=\frac{\int_0^\infty bdb \langle p_T\rangle (s,b) \langle n\rangle (s,b) h_{inel}(s,b)}
{\int_0^\infty bdb  \langle n\rangle (s,b)h_{inel}(s,b)}
\end{equation}
where $h_{inel}(s,b)$ is the inelastic overlap function 
Now calculating the respective integrals
 we obtain the power-like dependence
of the average transverse momentum $\langle p_T \rangle (s)$ at high energies
\begin{equation}\label{apt}
\langle p_T \rangle (s) = cs^{\delta_C},
\end{equation}
where
\[
 \delta_{C}={\frac{1}{2}\left(1-\frac{\xi }{m_QR_{C}}\right)}.
\]
The value of  $\xi$ is obtained from the data on angular
 distributions \cite{csn} and for the mass of constituent quark
 the standard value $m_Q=0.35$ GeV is taken. Of course, besides collective
effects average transverse momentum would get contributions from other sources
such as thermal distribution proposed long time ago by Hagedorn \cite{hag}.
This part has no energy dependence and we take it into account by simple addition of the constant
term to the power-dependent one, i.e.:
\begin{equation}\label{apte}
\langle p_T \rangle (s) = a+cs^{\delta_C}
\end{equation}
Existing experimental data can be described well (cf. Fig. 10) using Eq. (\ref{apte}) with parameters
$a=0.337$ GeV/c, $c=6.52\cdot 10^{-3}$ GeV/c and $\delta_C=0.207$.
\begin{figure}[h]
\begin{center}
\resizebox{8cm}{!}{\includegraphics*{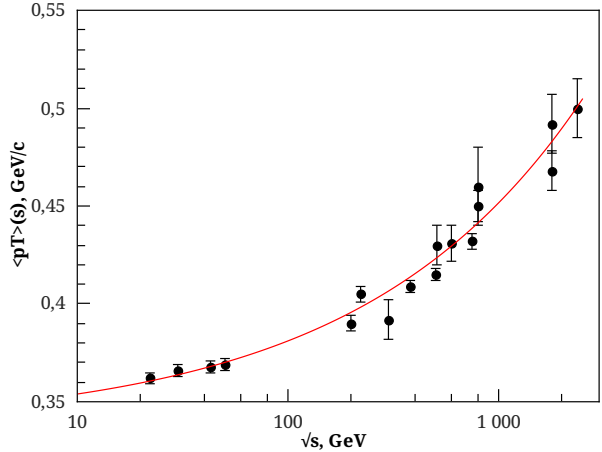}}
\end{center}
\caption{\small\it Energy dependence of the average transverse momentum in $pp$-collisions, experimental data from \cite{cmspt,all}.}
\end{figure}
The numerical value of $R_C$ is determined by pion mass with better than 10\% precision,
$ R_C\simeq 1/m_{\pi}$.
In the model the  indices in the energy dependencies of average multiplicity and transverse momentum $\delta$ and $\delta_C$ are
determined by the same expression and  experimental data fitting with free parameters $\delta$ and $\delta_C$ 
confirms this coincidence with better than 10\% precision also; note that the value $\delta=0.201$ follows from the experimental data analysis
for the average multiplicity.

 \section{Elliptic flow in proton collisions at the LHC energies}

As it was already noted the most typical inelastic event at the LHC energies occurs with nonzero impact parameter.
We consider therefore in this section peripheral hadronic collisions.   The orientation of the reaction plane in $pp$--collisions can
therefore be determined.

There are several experimental probes  of collective dynamics. Some of them were considered in the previous
sections. Another observables are related to the anisotropic flows and among them the  most widely discussed one
is the elliptic flows
\begin{equation}\label{v2}
v_2(p_T)\equiv \langle \cos(2\phi)\rangle_{p_T}=\langle \frac{p_x^2-p_y^2}{p_T^2}\rangle,
\end{equation}
which is the second Fourier moment of the azimuthal momentum distribution of the particles
 with a  fixed value of $p_T$.
The azimuthal angle $\phi$ is
an angle of the detected particle with respect to the reaction
plane, which is spanned by the collision axis $z$ and the impact parameter vector $\mathbf b$. The impact
parameter vector $\mathbf b$ is directed along the $x$ axis. Averaging is taken over a large number
 of the events. Elliptic flow can be expressed  in covariant form in
 terms of the impact parameter and transverse momentum
 correlations as follows
 \begin{equation}\label{v2a}
v_2(p_T)=\langle \frac{(\hat{\mathbf  b}\cdot {\mathbf  p}_T)^2}{p_T^2}\rangle-
\langle\frac{(\hat{\mathbf  b}\times {\mathbf  p}_T)^2}{p_T^2}\rangle ,
\end{equation}
where $\hat{\mathbf  b}\equiv \mathbf  b /b$. 
To get some hints on the possible behavior of the elliptic flow in proton collisions,
it is useful to recollect what is known on this observable from nuclear collisions experiments.
Integrated elliptic flow $v_2$  at high energies
is positive and increases with $\sqrt{s_{NN}}$.
The differential elliptic flow $v_2(p_T)$ increases with $p_T$
at small values of transverse
momenta, then it becomes flatten in the region of the intermediate transverse
momenta and decreases at large $p_T$. 
It is also useful  to apply a simple geometrical ideas which imply existence of the elliptic
flow in hadronic reactions.
Geometrical notions for description
of multiparticle production in hadronic reactions were used by many authors, e.g  by Chou and Yang in \cite{chyn}.
In the peripheral hadronic collisions the overlap region has different sizes along the $x$ and $y$ directions.
According to the uncertainty principle we can estimate the value
of $p_x$ as $1/\Delta x$ and correspondingly $p_y\sim 1/\Delta y $
where $\Delta x$ and $\Delta y$ characterize the size of the
region where the particle originate from. Taking $\Delta x \sim
R_x$ and $\Delta y \sim R_y$, where $R_x$ and $R_y$ characterize
the sizes of the almond-like overlap region in transverse plane,
 we can easily
obtain proportionality of $v_2$ in collisions with fixed initial impact 
parameter to the eccentricity of the overlap
region, i.e.
\begin{equation}\label{exc}
v_2\sim \frac{R_y^2-R_x^2}{R_x^2+R_y^2}.
\end{equation}

The presence of correlations of impact parameter vector $\mathbf
b$ and $\mathbf p_T$ in hadron interactions follows also from the relation between
impact parameters in the multiparticle production:
\begin{equation}\label{bi}
{\mathbf b}=\sum_i x_i{ {\mathbf  b}_i}.
\end{equation}
Here  $x_i$ stand for Feynman $x_F$ of $i$-th particle, the impact
parameters ${\mathbf b}_i$ are conjugated to the transverse
momenta ${\mathbf p}_{i,T}$. 

The above  considerations are  based on the uncertainty principle and angular momentum
conservation, but they do not preclude the existence of the dynamical description, which will be discussed
in the next section

As it was noted in the  section 6 an essential point of the proposed production mechanism is the 
rotation of the transient state due to the presence of a non-zero impact parameter in the collision.
The unitarity saturation  invokes a dynamical mechanism leading to the  peripheral nature
 of inelastic collisions at the LHC energies. 
One can suggest therefore that the events with average and higher multiplicity at the LHC energy $\sqrt{s}=7$ TeV
  correspond
to the peripheral hadron collisions.
Of course, the standard inclusive cross-section for unpolarized particles
being integrated over impact parameter $\mathbf b $,  does not depend on the
azimuthal angle of the detected
particle  transverse momentum. 

When the impact parameter vector $ \mathbf {b}$ and transverse momentum ${\mathbf  p}_T $
of the detected particle are fixed
the function $I$ does
  depend on the azimuthal angle $\phi$ between
 vectors $ \mathbf b$ and ${\mathbf  p}_T $.
The dependence on the azimuthal angle $\phi$ can be written in explicit form through the Fourier
series expansion
\begin{equation}\label{fr}
I(s,\mathbf b, y, {\mathbf  p}_T)=\frac{1}{2\pi}I_0(s,b,y,p_T)[1+
\sum_{n=1}^\infty 2\bar v_n(s,b,y,p_T)\cos n\phi].
\end{equation}
The function $I_0(s,b,\xi)$ satisfies  to the
following sum rule
\begin{equation}\label{sumrule}
\int I_0(s,b,y,p_T) p_T d p_T dy=\langle n \rangle (s,b)\mbox{Im} U(s,b),
\end{equation}
where $\langle n\rangle (s,b)$ is the mean multiplicity depending on impact parameter.
Thus, the bare anisotropic  flow $\bar v_n(s,b,y,p_T)$ is related to the
measured  flow $v_n$  as follows
\[
v_n(s,b,y,p_T)=w(s,b)\bar v_n(s,b,y,p_T).
\]
where the function $w(s,b)$ is
\[
w(s,b)\equiv |1-iU(s,b)|^{-2}.
\]
 In the above formulas the variable $y$ denotes rapidity, i.e. $y=\sinh^{-1}(p/m)$,
where $p$ is a longitudinal momentum.
Thus, we can see that unitarity corrections are mostly important
at small impact parameters, i.e. they modify anisotropic flows at small centralities,
while peripheral collisions are almost not affected by unitarity.

Now we are going to  use the particle production mechanism described above
for evaluation of the elliptic flow in $pp$-interactions.
It was suggested   that the rotation of transient matter
will affect the average transverse momentum of the secondary hadrons produced in proton-proton collisions.
Going further, one should be more specific and note that, in fact, the rotation gives a contribution to the $x$-component of the transverse 
momentum and does not contribute to the $y$-component of the transverse momentum, i.e.
\begin{equation} \label{ptx}
\Delta p_x=\kappa L(s,b)
\end{equation}
while
\begin{equation} \label{pty}
\Delta p_y=0.
\end{equation}
Assuming that $p_x=p_0+\Delta p_x$ ($\Delta p_x \ll p_0$) and $p_y=p_0$, 
the contribution  into the elliptic flow of the  transient matter rotation  can be calculated 
by analogy with the average transverse momentum calculation   performed in \cite{ptrans}.
The resulting  integral elliptic flow increases with energy
\[
v_2\propto s^{\delta_C} ,
\]
where  $\delta_C=0.207$. 
It should be noted here that transient matter consists of virtual constituent quarks strongly 
interacting
via pion (Goldstone bosons) exchanges .

The incoming constituent quark has a finite geometrical size determined by the radius $r_Q$ and 
interaction radius $R_Q$ ($R_Q>r_Q$). The former one is determined by the chiral symmetry 
breaking mechanism and the latter one --- by the confinement radius.
Meanwhile, it is natural to
suppose on the base of the uncertainty principle that size of the region where the virtual massive quark $Q$ is knocked out from the cloud
is determined by its transverse momentum, i.e. $\bar R\simeq 1/p_T$. However, it is
evident that $\bar R$ cannot be larger than the interaction radius of the valence
 constituent quark $R_Q$. 
It is also clear that $\bar R$ should not be less than the geometrical size of the valence constituent
quark $r_Q$ for this mechanism be a working one.  When  $\bar R$ becomes less than $r_Q$, 
this constituent quark mechanism does not work anymore
and one should expect vanishing collective effects in the relevant region of the transverse momentum.

The value of the quark interaction radius was
 obtained under analysis of the elastic scattering \cite{csn}
 and it has the following dependence on its mass
\begin{equation}\label{rq}
R_Q= \xi/m_Q \sim 1/m_\pi
\end{equation}
where $\xi \simeq 2$ and therefore $R_Q\simeq 1$ $fm$, while the geometrical radius of  quark $r_Q$
is about $0.2$ $fm$.
It should be noted  the region, which is responsible for the small-$p_T$ hadron production, has large transverse dimension and the incoming
constituent quark excites the rotating cloud of quarks with different values and directions
of their momenta in that case. Effect of rotation will therefore be smeared off  over the volume $V_{\bar R}$
 and then  one should expect that $\langle \Delta p_x \rangle_{V_{\bar R}} \simeq 0$. Thus,
\begin{equation}
\label{larg}
v^Q_2(p_T)\equiv\langle v_2\rangle_{V_{\bar R}}\simeq 0
\end{equation}
at small $p_T$.
When we proceed to the region of higher values of $p_T$, the radius $\bar R$ is decreasing
and the  effect of rotation  becomes more and more prominent, incoming valence quark excites now the region
where most of the quarks move coherently, in the same direction, with approximately
the same velocity. The mean value $\langle \Delta p_x \rangle_{V_{\bar R}} > 0$ and
\begin{equation}
\label{smal}
v^Q_2(p_T)\equiv\langle v^Q_2\rangle_{V_{\bar R}}> 0
\end{equation}
and it increases with $p_T$.
The increase of $v^Q_2$ with $p_T$ will disappear at
$\bar R =r_Q$, i.e. at $p_T \geq 1/r_Q$, and saturation will take place.
The value of transverse
momentum where the flattening starts is about $1$ $GeV/c$ for $r_Q\simeq 0.2$ $fm$. At very large transverse
momenta the constituent quark picture would not be valid and elliptic flow vanishes as it was already mentioned.

We discussed elliptic flow for the constituent quarks.
Predictions for the elliptic flow for the 
hadrons depends on the supposed mechanism of hadronization. For  the intermediate values of $p_T$ the constituent
quark coalescence mechanism \cite{volosh,volosh1} would be dominating one.
In that case  the hadron elliptic flow can be obtained from
the constituent quark one by the replacement $v_2\to n_Vv^Q_2$ and
$p_T\to p^Q_T/n_V$, where $n_V$ is the number of constituent quarks in the produced
hadron.

Typical qualitative dependence of elliptic flow in $pp$-collisions in this approach is presented in
Fig.11
\begin{figure}[h]
\begin{center}
  \resizebox{8cm}{!}{\includegraphics*{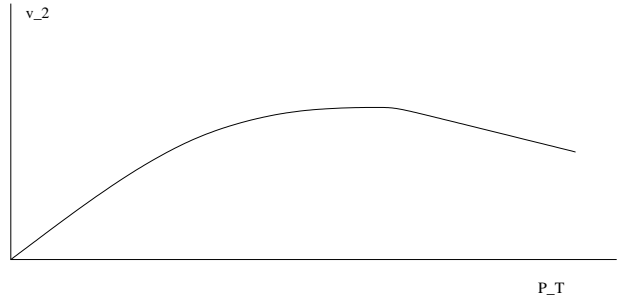}}
\end{center}
\caption{\small\it Qualitative dependence of the elliptic flow $v_2$ on transverse momentum in pp-collisions.}
\end{figure}

The centrality dependence of the elliptic flow is determined by
the orbital angular momentum 
  $L$  dependence on the impact parameter, i.e. it should   be decreasing towards  high and low centralities.
  Decrease toward high centralities is evident since no overlap of hadrons should occur at high enough
  impact parameters. Decrease of $v_2$ toward lower centralities is specific prediction of the proposed
  mechanism based on rotation
  since the central collisions with smaller impact parameters would lead to slower rotation or its complete
   absence in the head-on collisions (it is also result of the symmetry in head-on collisions). 
Qualitative dependence of the elliptic flow on the impact parameter
  is similar to the one depicted in Fig.11 where variable $b$ is to be used instead of transverse
 momentum.
\section{Two-particle correlations in $pp$-collisions}

The ridge structure was observed first at RHIC  in the two-particle 
correlation function in the near-side
jet production. It was demonstrated that the ridge particles have
 a narrow $\Delta\phi$ correlation distribution (where $\phi$ 
is an azimuthal angle) and wide $\Delta\eta$ correlations ($\eta$ is a pseudorapidity). The ridge phenomenon was associated
with the collective effects of a medium \cite{rhicr}. 

The similar structure in the two-particle correlation function was observed by
the CMS Collaboration \cite{ridgecms}. This is rather surprising result because the ridge structure was observed for the first time
in $pp$--collisions. Those collisions are commonly treated as  the ``elementary'' ones under the heavy-ion studies 
and therefore often used  as the reference
process for detecting deconfined phase formation on the base of difference between $pp$- and $AA$-collisions. 
It is evident now that such approach should be revised in view of this new and  
unexpected  experimental result.

In the proposed explanation of the ridge effect at the LHC energy $\sqrt{s}=7$ TeV  a dynamical
selection of peripheral region in impact parameter space responsible for the inelastic processes is the important point.
As it was already noted, the  geometrical picture of hadron collision at non-zero impact parameters
implies that the generated massive
virtual  quarks in overlap region  could obtain very large initial orbital angular momentum
at high energies. Due to strong interaction
between the quarks this orbital angular momentum  leads to the coherent rotation
of the quark system located in the overlap region as a whole  in the
$xz$-plane. This rotation is similar to the rotation of  liquid. 
The assumed particle production mechanism at moderate transverse
momenta is an excitation of  a part of the rotating transient state of  massive constituent
quarks (interacting by pion exchanges). 
\begin{figure}[h]
\begin{center}
\resizebox{7cm}{!}{\includegraphics*{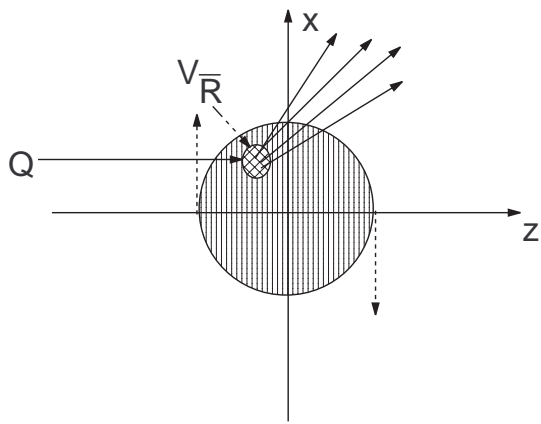}}
\caption{\small \it {Interaction of the constituent quark with rotating quark-pion liquid.}}
\end{center}
\end{figure}
Due to the fact that the transient matter is strongly interacting, the excited parts
should be located closely  to the periphery  of the rotating transient state otherwise absorption
 would not allow to quarks and pions  leave the interaction region (quenching). 

The mechanism is sensitive
 to the particular  direction of rotation and to the rotation plane orientatation. This will
lead to the narrow distribution of the two-particle correlations in $\Delta\phi$. However,
two-particle correlation could have broad distribution in polar angle ($\Delta\eta$) in the above mechanism (Fig. 12).
 Quarks in the exited part of the cloud
could have different values of the two components of the momentum (with its third component
lying in the rotation  $xz$-plane) since the exited region $V_{\bar R}$ has significant extension. 

Thus,  the ridge-like structure observed  in the high multiplicity events by the CMS Collaboration 
can be considered as  experimental manifestation of the coherent rotation of the transient matter in hadron
collisions. The narrow  two-particle correlation distribution in the azimuthal angle is the
distinctive feature of this mechanism. 

There should be other experimental
observations of this collective effect.  One  was mentioned already.  It is the directed flow $v_1$ in hadron reactions,
with fixed impact parameter discussed in \cite{qgpo}.  Rotation of transient matter
will affect also elliptic flow  $v_2$ and  average transverse momentum of the secondary particles produced in proton-proton collisions.

The ridge effect has triggered flow of  possible explanation using different mechanisms \cite{Shur,ColGl,Bozek,Drem,Train,Cher,Kovner,Wer,ADrem,Hwa,Savrin}.

The  one described in this paper and in \cite{ridge} relates appearance of the ridge effect with  
 the reflective scattering at the LHC energies.
Absence of the reflective scattering at  lower energies explains why the ridge effect was not observed at the ISR or 
Tevatron. This important point on the ridge energy dependence needs to be explained by any particular mechanism 
proposed as a source of the two-particle correlations.

\section{Genuine QGP  formation}
The dynamical mechanism of the average transverse momentum growth, anisotropic flows $v_1$, $v_2$,
 two-particle correlations (ridge) originate from the collective effect
of transient matter rotation, while dynamics of average multiplicity growth is related to the  mechanism where a nonlinear 
field couplings   transform  the kinetic energy to
internal energy.  
Formation of a genuine quark-gluon plasma in transient state in the form of the noninteracting gas of free quarks and gluons
would result in disappearance of the effects of rotation of the transient state. 
 Vanishing  the energy dependent
contribution to the average transverse momentum should therefore be expected, i.e. average transverse momentum would reach maximum
at some energy and then will decrease till eventual flat behavior (cf. Fig.13). 
\begin{figure}[hbt]
\begin{center}
\resizebox{10cm}{!}{\includegraphics*{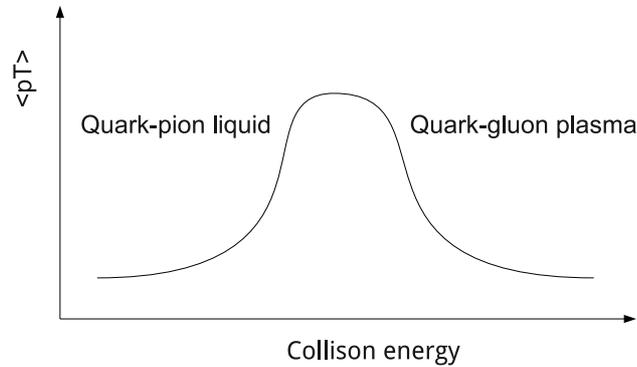}}
\end{center}
\vspace{-1cm}
\caption{\small\it Qualitative energy dependence of the average transverse momentum in $pp$-collisions  in case of genuine QGP formation at
super high energies.}
\end{figure}
Vanishing rotation will lead to vanishing anisotropic flows $v_1$ and $v_2$.
Of course, this picture is a rather qualitative one since
the nature of phase transition from strongly interacting matter (associated with quark-pion liquid in the model) 
to the genuine quark-gluon plasma is not known. Currently, there is no  experimental indications
that such a transition will indeed take place  at super high energies. At the same time there are theoretical arguments
that existence of genuine QGP (ideal gas) would contradict to the confinement prop[erty of QCD. This contradiction
arises in the many dynamical mechanisms of hadronization \cite{mie,mel}. However, the issue of the genuine
QGP  existence can  only be resolved by the experimental searches.

 It should be  noted that inverse phase transition (from parton
gas in parton model to liquid) could explain a saturation phenomena in deep inelastic processes \cite{jenk}. 

The natural question appears  about the role of orbital angular momentum in those mentioned above
 phase transitions.
\section{Spin correlations due to reflective scattering in the case of genuine QGP formation}
It is interesting to make a conclusion on the possible effect of the imbalance of the 
orbital angular momentum in the case of the genuine QGP formation.
It should be noted that we can consider separately particles production in the forward and
backward hemispheres \cite{webber}. Let us consider for example
particles produced in the forward hemisphere. The  orbital angular
momentum in the initial state can be estimated as
\begin{equation}\label{li}
L_i\simeq \frac{\sqrt s}{2}\frac{R(s)}{2},
\end{equation}
where $R(s)$ is the interaction radius. The orbital angular
momentum in the final state is then
\begin{equation}\label{lfi}
L_f\simeq \frac{\langle n \rangle  (s)\langle x_L \rangle (s)\sqrt s}{4}\frac{R(s)}{2},
\end{equation}
where we have taken into account that $\langle n \rangle (s)/2$ particles
with the average fraction of their longitudinal momentum
$\langle x_L \rangle (s)$ are produced at the impact parameter $R(s)/2$ due to
reflective scattering. The average fraction of longitudinal momentum
$\langle x_L\rangle (s)$  according to the hypothesis
 of limiting fragmentation \cite{yen} would not decrease with energy. Thus
 we arrive to the negative imbalance of the orbital angular momentum
\begin{equation}\label{lf}
\Delta L=L_i-L_f\simeq \frac{\sqrt s}{2}\frac{R(s)}{2}(1- \frac{\langle n \rangle (s)\langle x_L\rangle (s)}{2}),
\end{equation}
i.e.
\begin{equation}\label{lfd}
\Delta L \simeq -\frac{\sqrt s}{2}\frac{R(s)}{2}\frac{\langle n\rangle  (s)\langle x_L\rangle (s)}{2}.
\end{equation}
This negative $\Delta L$ should be compensated by
 the total positive spin $S$ of final particles (since the particles in the initial state
 are unpolarized)
\begin{equation}\label{sp}
S=-\Delta L
\end{equation}
This  spin alignment of produced particles appears in the
reflective scattering mode when the particles are produced in the region
of impact parameters $b=R(s)$.

The vector of spin $\vec{S}$ lies in the transverse plane but it
cannot be detected through the transverse polarization of single
particle due to azimuthal symmetry of the production process
(integration over azimuthal angle $\varphi$). However, this effect
can be traced measuring transverse spin correlations of two
particles $\langle s_i s_j\rangle $.  The most evident way to
reveal this effect is to perform the measurements of the spin
correlations of hyperons whose polarizations can be extracted from
the angular distributions of their weak decay products.

Spin correlation should be stronger for the light particles and the
weakening is expected for heavy particles since they should be
produced at smaller values of impact parameters.

Thus,  we can expect appearance at the LHC energies of strong spin
correlations of final particles as a result of the prominent
reflective scattering mode in the case if genuine
 QGP is formed in the transient state. Vanishing anisotropic flows, decreasing average
transverse momentum and
appearance (simultaneous) of the secondary particles polarization are
thus the signals of the genuine QGP formation.

\section{Effects of the reflective scattering mode for nuclear collisions}
In this final section we proceed from proton collisions to the collisions of nuclei.
Consider central collision of two identical nuclei having $N$ nucleons in total with center of mass energy
$\sqrt{s}$ per nucleon and calculate
nucleon density $n_R(T,\mu)=N/V$ in the initial state at given
 temperature $T$ and baryochemical potential $\mu$ in the presence of
the reflective scattering.
The effect of the reflective scattering of
  hadrons is equivalent  to decrease of the volume of the available  space
  which the hadrons are able to occupy in the case when reflective scattering is absent.
  Thus followings to van der Waals method, we must then replace volume $V$ by $V-p_R(s)V_R(s)\frac{N}{2}$,
  i.e. we should write
  \[
  n(T,\mu)=\frac{N}{V-p_R(s)V_R(s)\frac{N}{2}},
\]
where $n(T,\mu)$ is hadron density  without account for reflective scattering and
 $p_R(s)$ is the averaged over volume $V_R(s)$ probability of reflective scattering:
\[
p_R(s)=\frac{1}{V_R(s)}\int_{V_R(s)}|S(s,r)|^2 d^3x.
\]
The volume $V_R(s)$ is determined by the radius of the reflective scattering.
Here we  assume spherical symmetry of hadron interactions,
i.e. we replace impact parameter $b$ by $r$ and approximate the volume $V_R(s)$ by
$V_R(s)\simeq (4\pi/3)R^3(s)$. Hence, the density $n_R(T,\mu)$  is connected
with corresponding density in the approach without reflective scattering $n(T,\mu)$
 by the following relation
\[
  n_R(T,\mu)=\frac{n(T,\mu)}{1+\alpha(s)n(T,\mu)},
\]
where $\alpha(s)={p_R(s)V_R(s)}/{2}$. Let us now estimate change of the function
$n_R(T,\mu)$ due to the presence of reflective scattering. We can
approximate $p_R(s)$ by the value of $|S(s,b=0)|^2$ which tends to
unity at $s\to\infty$. It should be noted that the value
$\sqrt{s_R}\simeq 2$ $TeV$ \cite{pras}. Below this energy there is
no reflective scattering, $\alpha(s)=0$ at $s\leq s_R$, and
therefore corrections to the hadron density are absent. Those
corrections are small when the energy is not too much higher than
$s_R$. At $s\geq s_R$ the value of $\alpha(s)$ is positive,
and presence of reflective
scattering diminishes hadron density.
\begin{figure}[hbt]
\begin{center}
\includegraphics[scale=0.3]{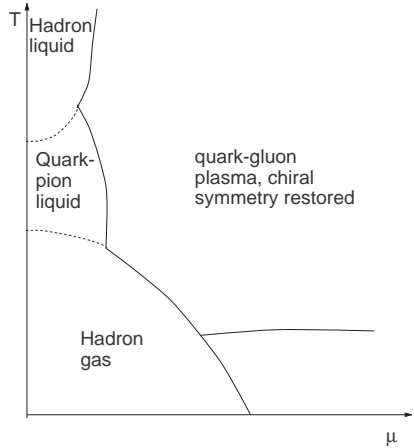}
\caption{\small\it Phases of strongly interacting matter.}
\end{center}
\end{figure}
 We should expect that this
effect would already be  noticeable at the LHC energy $\sqrt{s}\simeq 5$
TeV in $Pb+Pb$ collisions. At very high energies ($s\to\infty$)
\[
n_R(T,\mu)\sim 1/\alpha(s)\sim M^3/\ln ^3 s.
\]
This  limiting dependence for the hadron density  appears due to
 the presence of the reflective scattering which results in similarity of head-on hadron collisions
with  scattering of hard spheres.  It should be noted
that this dependence has been obtained under assumption on spherical symmetry of hadron interaction region.
Otherwise, limiting dependence of the hadron density in transverse plane can only be obtained,
i.e. transverse plane density of hadrons would have then the following behavior
\[
n_R(T,\mu)\sim M^2/\ln ^2 s.
\]

Thus, the following chain of  transitions between different (physical and nonperturbative) vacuum states can be foreseen 
as the temperature increases at the constant value of chemical
potential $\mu$:
\[
 |0\rangle_{ph}(\mbox{Hadron gas})\to |0\rangle_{np}(\mbox{Quark-pion liquid})\to |0\rangle_{ph}(\mbox{Hadron liquid}).
\]
The corresponding phase diagram depicted in Fig. 14.

Thus, the reflective scattering mode 
can help in searches of the
deconfined state and studies  of transition mechanism to this state of matter. 
\section*{Acknowledgement}
We had pleasure discussing results described in this paper  with
N.~Buttimore, J.~Cudell,  M.~Islam,  L.~Jenkovszky,  A.~Krisch, E.~Levin, I.~Lokhtin, U.~Maor, A.~Martin, 
E.~Martynov,  V.~Petrov,  A.~Prokudin, J.~Ralston,  V.~Savrin, D.~Sivers and O.~Teryaev.

\end{document}